\documentclass[3p,times]{elsarticle}

\usepackage{ecrc}

\usepackage{epstopdf}

\volume{00}

\firstpage{1}

\journalname{Nuclear Physics A}

\runauth{Sayantan Sharma}

\jid{nupha}

\jnltitlelogo{Nuclear Physics A}

\usepackage{graphicx}
\usepackage{amsmath,amssymb}

\begin{document}

\begin{frontmatter}



\title{The thermodynamics of heavy light hadrons at freezeout}

\author{Sayantan Sharma (for the Bielefeld-BNL-CCNU\fnref{col1} Collaboration)}
\fntext[col1] {A. Bazavov, H.-T. Ding, P. Hegde, O. Kaczmarek, F. Karsch,
E. Laermann, Y. Maezawa, S. Mukherjee, H. Ohno, P. Petreczky,
C. Schmidt, S. Sharma, W. Soeldner and M. Wagner}
\address{ Fakult\"{a}t f\"{u}r Physik, Universit\"{a}t Bielefeld, D-33615 Bielefeld, Germany}




\begin{abstract}
In the discussion of hadronization at or close to the freeze-out curve statistical (hadron resonance gas) 
models play an important role. In particular, in the charmonium sector, regeneration models are 
considered which rely on the fact that charmonium states can form again already at temperatures 
well above the QCD transition or hadronization temperature. An important ingredient in these 
considerations is the regeneration or hadronization of open charm states. In this talk we report 
on a lattice QCD analysis of correlations of open strange and charm with other conserved 
quantum numbers like the net baryon number and electric charge. We analyze the temperature range 
in which an uncorrelated hadron resonance gas (HRG) provides an adequate description of such 
correlations. This limits the range of validity of HRG based thermodynamics in open flavor 
channels and provides an estimate for the melting temperature of heavy-light hadrons. 
\end{abstract}


\end{frontmatter}



\section{Introduction}
\label{intro}
The strongly interacting matter described with Quantum Chromodynamics shows many 
interesting features like deconfinement and chiral symmetry breaking. For 
QCD with light quark flavours in the fundamental representation, the chiral 
symmetry restoration and deconfinement of the light quark degrees of freedom 
occur at the same temperature which is known as the chiral crossover temperature $T_c$. 
It is interesting to understand whether the same  phenomenon occurs for the heavier 
strange and charm quarks. It is known that the bound states of charm quarks and its 
anti-particle like $J/\psi$, $\eta_c$ deconfine at temperatures about $1.5~T_c$~\cite{ding,petreczky} and 
serve as a probe of the presence of a quark gluon plasma~\cite{satz}. Recently it has been shown 
from first principles lattice study that the open strange hadrons melt at $T_c$~\cite{strangemelt}, 
confirmed through another independent analysis~\cite{strangebw}. The 
charm quark mass is an order of magnitude larger than the strange mass and it is important 
to understand when the open charm hadron states do melt. Apart from being a problem of 
fundamental importance, it has phenomenological implications too. At LHC energies, it is 
expected that the charmonium states would be regenerated at the freezeout temperatures~\cite{PBM}. In 
the statistical model of hadronization inspired by the hadron resonance gas(HRG) model, the 
open charm hadrons would contribute to the regeneration of the charmonium states through feed 
down corrections. For these studies it is important to know the hadron content at the freezeout. 
In this talk we address two issues pertaining to the heavy light hadrons 
especially in the charm sector. From first principles lattice study we estimate 
the melting temperature for the open charm mesons and baryons independently of the 
hadron content. We then proceed to show that QCD thermodynamics at the chiral crossover transition 
provides hints about the existence of many more charmed baryons which are yet to be detected 
in the experiments.

\section{Melting of open charm hadrons}
\label{melt}

The total pressure of a thermal ensemble of a non-interacting gas of open charm mesons and baryons 
and resonances at finite chemical potential $\hat\mu_X=\mu_X/T~,X\equiv B,C$ is given as,
\begin{equation}
 P(\hat\mu_C,\hat\mu_B)=P_M\cosh (\hat\mu_C)+P_{B,C=1}\cosh(\hat\mu_B+\hat \mu_C)+ 
 P_{B,C=2}\cosh(\hat\mu_B+2\hat \mu_C) +P_{B,C=3}\cosh(\hat\mu_B+3\hat \mu_C)~.
\end{equation}
Motivated from the analysis developed in Ref.~\cite{strangemelt}, we can rewrite the partial 
pressures in the baryon and the meson sectors in terms of susceptibilities at $\hat\mu_X=0$.
The derivatives of $P(\hat\mu_C,\hat\mu_B)$ with respect to $\mu_X$, $\chi_{kl}^{BC} = \left. 
\frac{\partial^{(k+l)} [P(\hat\mu_B,\hat\mu_C)/T^4]} {\partial \hat\mu_B^k \partial \hat\mu_C^l} \right|_{\vec{\mu}=0}$, 
yield different susceptibilities which denote the fluctuations of $B,C$ and their correlations. 
In a non-interacting gas of charmed hadrons, these quantities can be written as,
\begin{equation}
\chi_n^C(\hat\mu=0)=P_M+1^n P_{B,C=1}+2^n P_{B,C=2}+ 3^n P_{B,C=3}~,~
\chi_{kl}^{BC}(\hat\mu=0)=1^l P_{B,C=1}+2^l P_{B,C=2}+ 3^l P_{B,C=3}~.
\end{equation}
Inverting these relations would give us the partial pressures in the meson and baryon sectors.
We calculate correlations between the conserved quantum numbers, net baryon number, charm and their
fluctuations upto fourth order at $\mu_X=0$ on the lattice on $2+1$ flavour Highly Improved Staggered 
Quark(HISQ) configurations with quenched charm quark. The strange quark mass is set to its 
physical value and the light quark mass is chosen such that $m_\pi=160$ MeV in the continuum. The charm quark 
mass is fixed by setting the spin-averaged charmonium mass \cite{Yu}, $\frac{1}{4} (
m_{\eta_c} + 3 m_{J/\psi})$ to its physical value. Two different lattices with spatial size 
$N_s=32,24$ and temporal extent $N_\tau=8,6$ respectively are used for this study. The crossover temperature 
is $T_c=154(9)$ MeV~\cite{bazavov} in the continuum limit for physical quarks. We analyzed 
susceptibilities in the temperature range $156$-$330$ MeV.
To control systematic errors, 6000 stochastic sources were employed to compute the fluctuations 
in the charm sector and 1500 for the light sector. About 16000 configurations were considered at the 
lowest temperature and 1600-6000 configurations at high temperatures to control the statistical errors. 
The computational details are outlined in ~\cite{charm}.

 \begin{figure}
\begin{center}
\includegraphics*[scale=0.45]{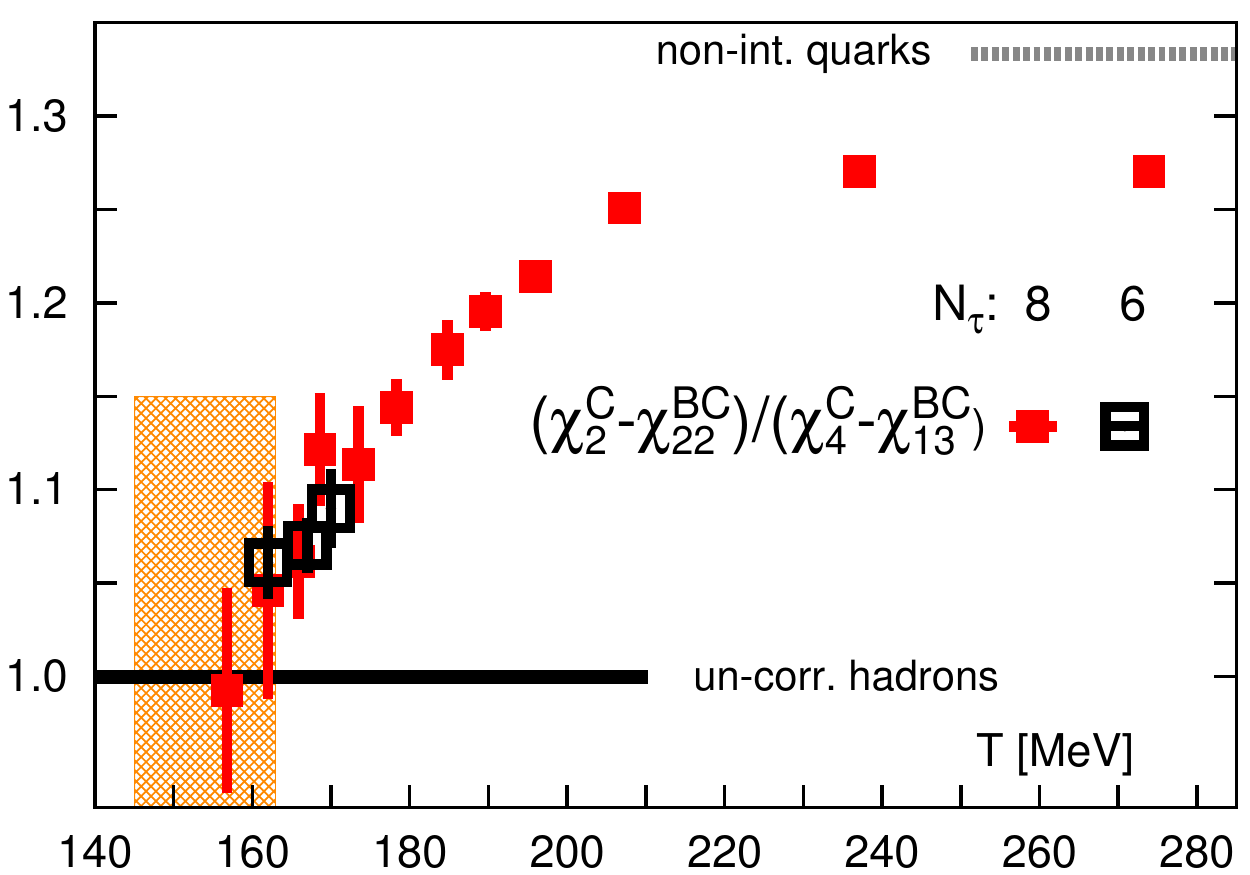}
\includegraphics*[scale=0.45]{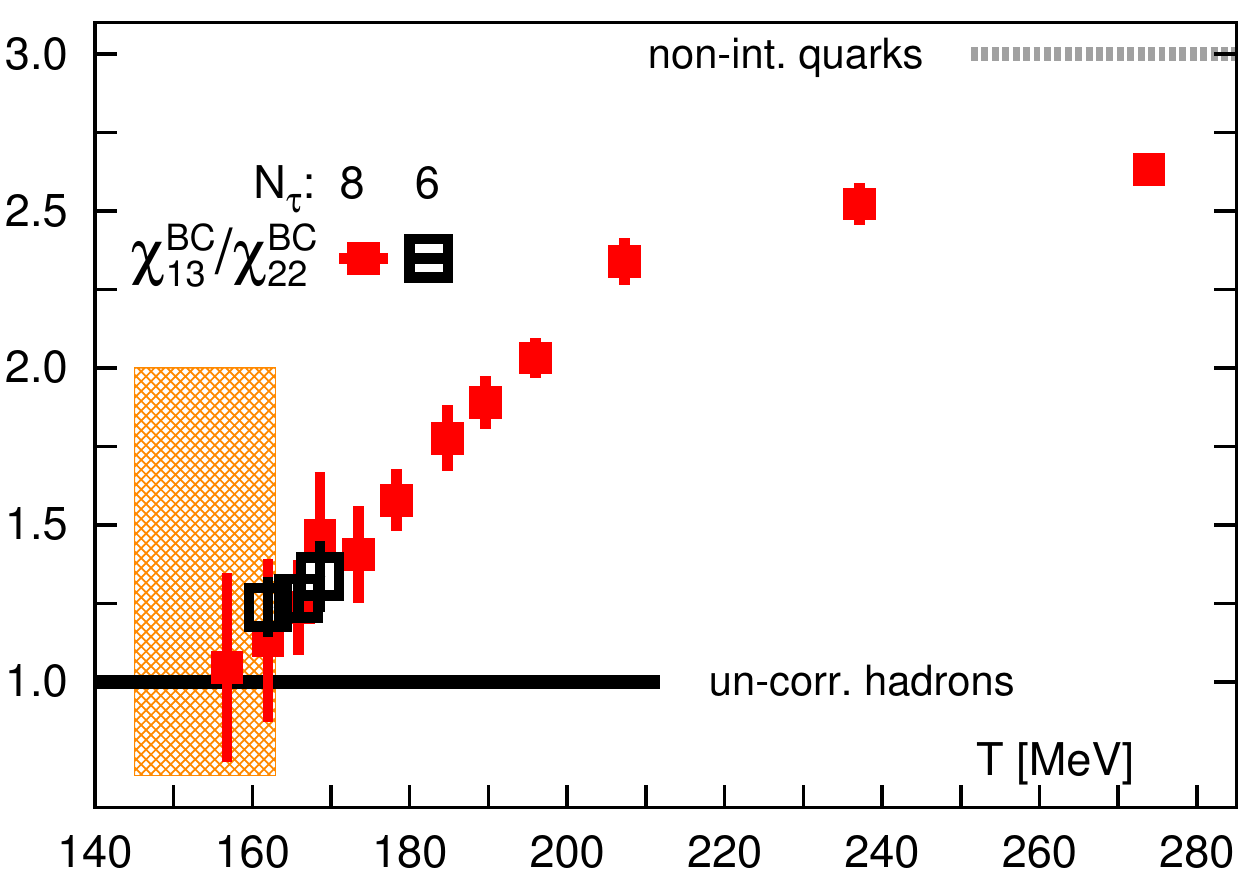}
\caption{The ratios of the partial pressure of the open charm mesons(left panel) and baryons(right panel) are 
shown as function of temperature for two different lattice sizes with $N_\tau=6,8$. The yellow band represents the 
crossover region. }
\label{fig:melting1}
\end{center}
\end{figure}

\begin{figure}
\begin{center}
\includegraphics*[scale=0.45]{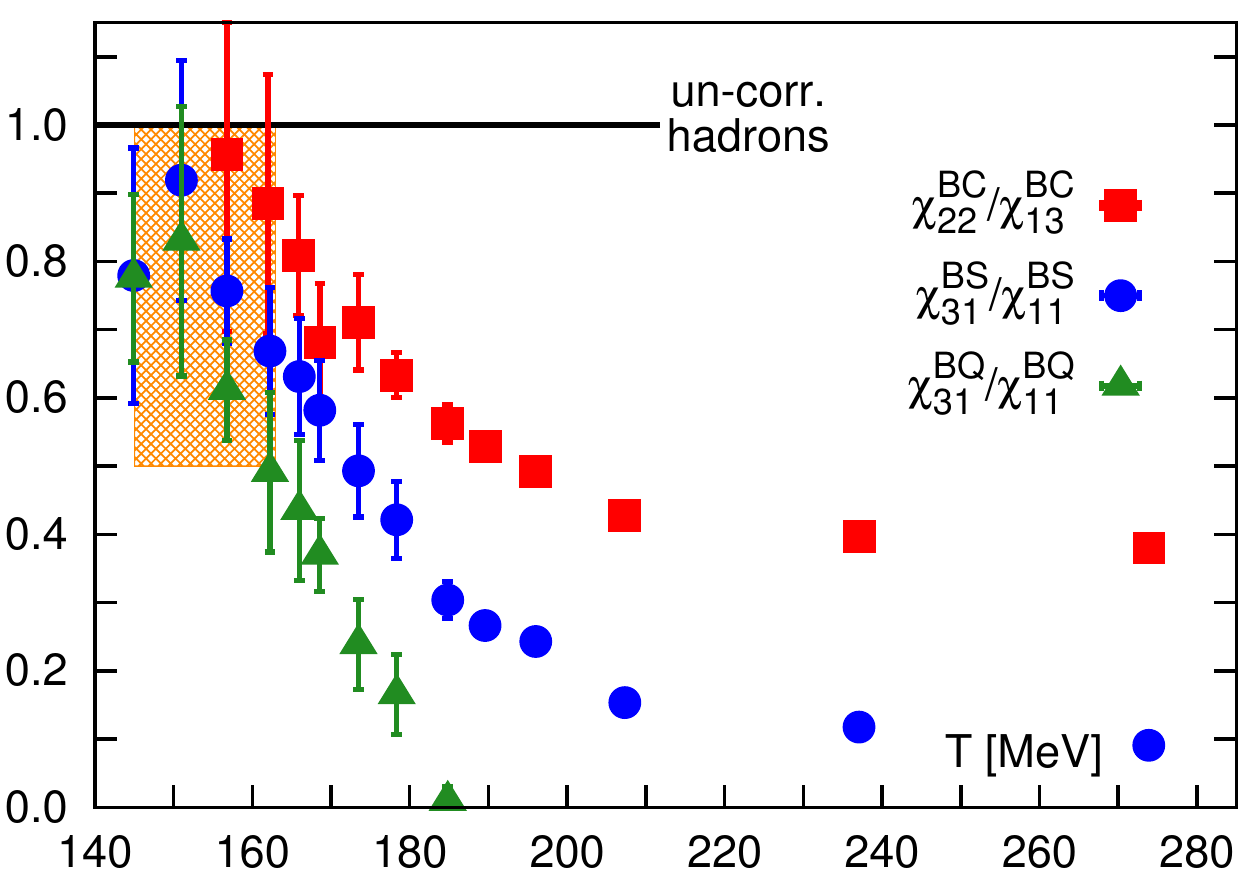}
\caption{The ratios of susceptibilities which probe the charm(BC) baryon, strange baryon(BS) and dominantly 
the light baryons(BQ) all show a departure from unity at the crossover region(shaded yellow).  }
\label{fig:melting2}
\end{center}
\end{figure}

The $P_M$ can be expressed as two equivalent combinations of the susceptibilities,
$P_M=\chi_2^C-\chi_{22}^{BC}=\chi_4^C-\chi_{13}^{BC}$.
The ratio of the two independent measures of $P_M$ would be unity in the hadron phase and deviation 
from unity would be a signal of the melting of the open charm mesons. In the baryon sector, $C=2,3$ baryons 
are substantially heavier than the $C=1$ baryons and their contribution to the total pressure has been shown 
to be negligibly small compared to $P_{B,C=1}$~\cite{charm}. It follows that $ P_{B=1}\simeq \chi_{mn}^{BC}~,m,n>0$ 
and $m+n=$ even. Ratios like $\chi_{13}^{BC}/\chi_{22}^{BC}$ would be unity in a phase of charmed baryons. 
The ratios in the charmed meson and baryon sectors are shown in Fig.~\ref{fig:melting1}. It is evident that both the open charm mesons 
and baryons melt at the crossover. These ratios are independent of the details of the hadron spectrum
and have relatively small lattice cut-off effects evident from the $N_\tau$ independence of our results. 
The behaviour of the light, strange and the charm baryons near the crossover are summarized in 
Fig.~\ref{fig:melting2} through the ratios of fluctuations which are chosen such that these are 
unity in the hadron phase. The lattice data in these sectors supports that deconfinement of the open 
heavy hadrons occur at the crossover transition region evident from the rapid departure of unity beyond 
$\sim 160$ MeV.

\section{Evidence of additional charm hadrons from QCD thermodynamics}
\label{charmhrg}
\begin{figure}
\begin{center}
\includegraphics*[scale=0.45]{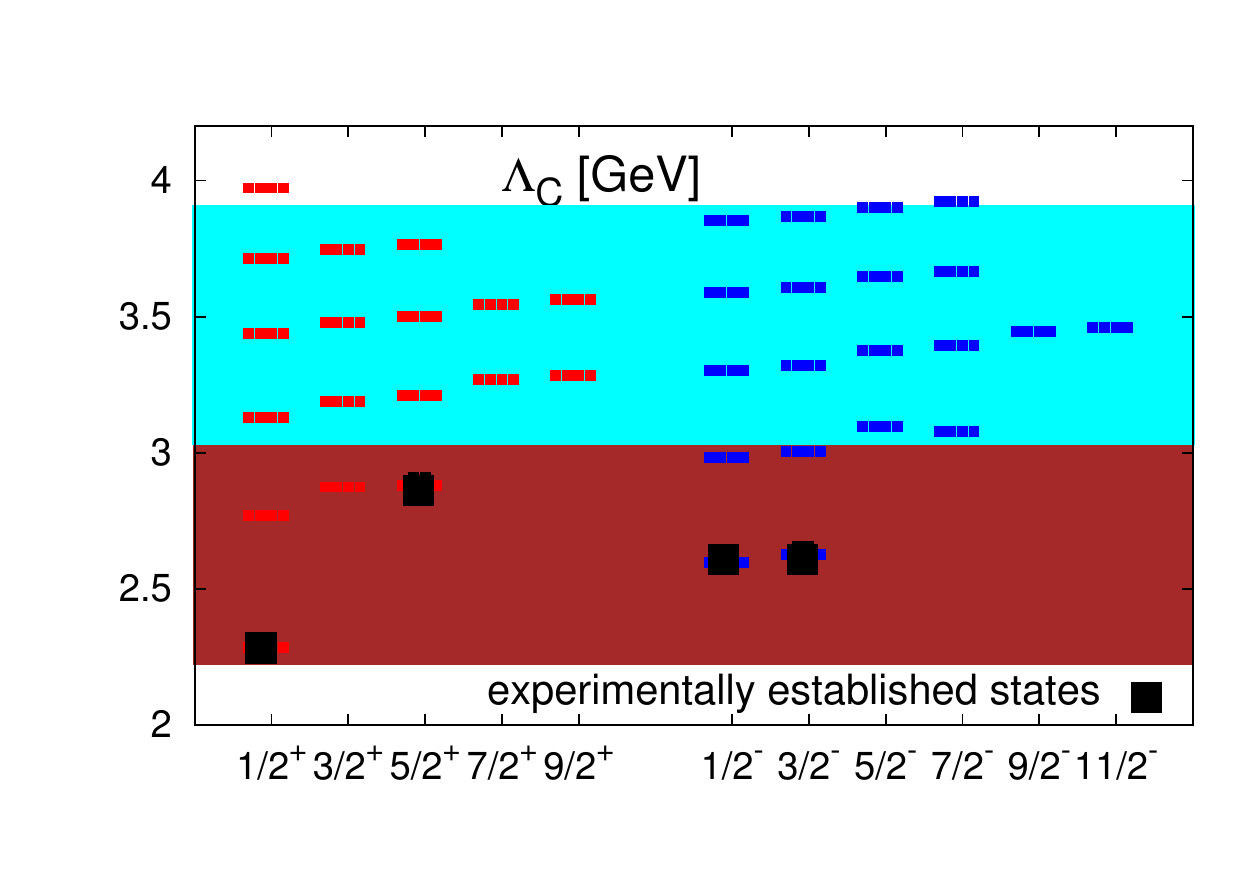}
\includegraphics*[scale=0.35]{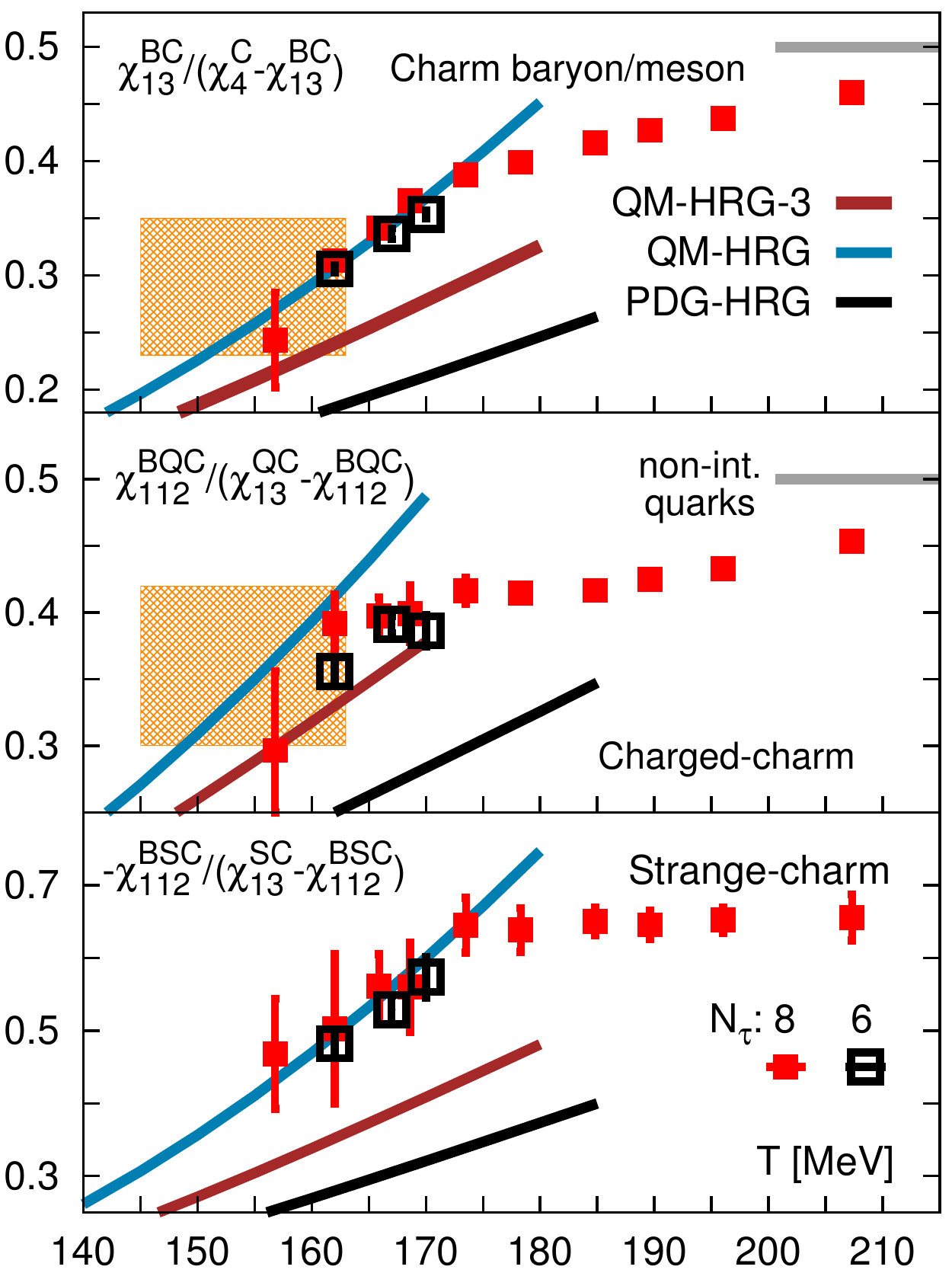}
\caption{The spectrum of charm Lambda baryons from the Quark Model calculations in the left panel has many more states 
that are not yet detected in the experiments. The brown band represents open charm hadrons with masses less than $3$ GeV 
and the blue band represents all such known states. In the right panel, our lattice QCD data of the ratio of partial 
pressures of the open charm mesons and baryons supports  the existence of the additional charm baryon states in 
different channels which are predicted from the Quark Model and hadron spectrum studies on the lattice.}
\label{fig:charmhrg}
\end{center}
\end{figure}

Our analysis allows us to disentangle the partial pressures of the meson and baryon sectors 
with different quantum numbers and understand the composition of these individual sectors at the 
crossover transition. Calculations in the Quark Model(QM)~\cite{Isgur,cQM,Ebert} as well hadron 
spectrum studies on the lattice~\cite{Edwards} predict many more open charm baryons 
than experimentally detected and tabulated in the Particle Data Table~\cite{pdg}, whereas the 
charmed meson sector is in good agreement~\cite{Ebertm,Moir}. For instance excitations which are 
$700$ MeV above the ground state are known for the $D,D_s$ mesons. On the other hand, the presence of the 
additional baryon states and resonances would have an impact on the pressure at the freezeout. 
In the left panel of Fig.~\ref{fig:charmhrg} 
we look at a particular charmed baryon $\Lambda_c$. Only three states above the ground state are measured in the 
experiments with well defined spin and parity and the ground state spin is also not measured with certainty 
yet. There are many more bound states in different spin-parity channels predicted from the QM. 
The lattice spectrum calculations yield very similar results. Many of the baryon states which have 
masses $1$ GeV or more than the ground state above the blue band in the figure may be unstable 
or would have negligible contribution to the partial pressures. To have a systematic understanding of the 
relative contribution of these states we compare with our lattice data, the pressure due to a hadron resonance gas(HRG) with all predicted 
states in the Quark Model (QM-HRG)~\cite{Ebert} denoted by brown lines, a QM-HRG-3 consisting of Quark Model states below $3$ GeV mass 
and a PDG-HRG~\cite{pdg} consisting of all the experimentally known states. We consider the partial 
pressure of charmed baryons with $C=1$ normalized by the partial pressure of the corresponding meson sector introduced in 
the earlier section, since ratios 
are not affected by lattice cut-off effects. Using appropriate susceptibilities, we also study the partial pressures in 
sectors with specific quantum numbers. In the charged sector, charmed baryon and meson partial pressures are given as $P_B\simeq\chi_{112}^{BQC}$  and $P_M\simeq\chi_{13}^{QC}-\chi_{112}^{BQC}$ whereas for the charmed baryons with strangeness $P_B\simeq\chi_{112}^{BSC}$  and $P_M\simeq\chi_{13}^{SC}-\chi_{112}^{BSC}$. 
The results are shown in the right panel of Fig.~\ref{fig:charmhrg}. 
In all these sectors we find a significant contribution of the additional baryon states over PDG-HRG at the 
crossover transition and our lattice data provides support in favour of the existence of these additional states.

\section{Conclusions}
We addressed the issue about the melting of hadron states consisting of light and strange quarks or 
the charm quarks. From our analysis of the correlations 
between the net baryon number and charm and their fluctuations on the lattice, we are able 
to disentangle the partial pressures of the meson and baryon sectors with different charm content.
The open charm mesons and baryons are shown to melt at the chiral crossover transition irrespective 
of the details of the hadron spectrum and lattice cut-off effects. 
We are for the first time able to extract the partial pressures of charmed baryon channel and even study 
the charmed baryons with different charge and strangeness content in QCD. This gives us a tool to understand 
the details of the charmed hadron spectrum in the crossover transition region. Quark Model as well as 
hadron spectrum studies on the lattice have reported the presence of many more charmed baryon states 
than those which have been measured in the experiments. These states would contribute to the thermodynamic 
quantities near the freezeout. Our data provides hints in support of the existence of these additional 
baryon states. These additional baryons would 
contribute to the hadronization of the quark gluon plasma believed to be formed in the 
heavy ion collision experiments and influence the abundances of hadrons through feed down 
corrections. Heavy-light hadrons also play a role in the dissociation of charmonium states and 
these additional states should be taken into account in all such considerations. A similar 
analysis has been recently performed for the strange sector as well~\cite{bibnlstrange} which also 
leads to evidence for the presence of additional strange baryons and important consequences like 
a lower strangeness freezeout temperature.

\section{Acknowledgements}
This work has been supported in part through contract DE-AC02-98CH10886 with the U.S.
Department of Energy, through Scientific Discovery through Advanced Computing
(SciDAC) program funded by U.S. Department of Energy, Office of Science, Advanced
Scientific Computing Research and Nuclear Physics, the BMBF under grant 05P12PBCTA, 
EU under grant 283286 and the GSI BILAER grant.
Numerical calculations have been performed using GPU-clusters at JLab, Bielefeld
University, Paderborn University, and Indiana University. We acknowledge the support
of Nvidia through the CUDA research center at Bielefeld University.








\end{document}